\documentclass[twocolumn]{elsart}

\usepackage{natbib}

\usepackage{epsfig}

\usepackage{amssymb}

\begin{document}

\begin{frontmatter}



\title{The Radio Jet Velocities at High Resolution}


\author{Gabriele Giovannini}

\address{Dipartimento di Astronomia, Universita' di Bologna, via Ranzani 1,
40127 Bologna and Istituto di Radioastronomia del CNR, via Gobetti
101, 40129 Bologna (Italy)}

\ead{ggiovann@ira.cnr.it}

\begin{abstract}
The different methods to derive the jet velocity and orientation
on the parsec 
scale are presented. From these methods I will discuss the velocity
distribution of parsec scale jets and the possible presence of acceleration or
deceleration in the jet velocity. Moreover, evidences of jet 
velocity structures
will be reported. Finally 
I will present new data on the superluminal giant radio source 1144+35
to discuss in detail the properties of pc scale jets.
\end{abstract}

\begin{keyword}
AGN, radio jets, jet velocity


\end{keyword}

\end{frontmatter}

\section{Introduction}
\label{}

Jets are the structures through which energy and matter can flow from the
Active Galactic Nucleus (AGN) to the extended radio lobes. 
There are now widely accepted lines of evidences for the
existence of relativistic bulk velocities in the
parsec scale jets of radio sources. This result implies that the observational
data are affected by relativistic effects (Doppler boosting) and 
provides the basis of
the current {\it unified theories} (see e.g. \cite{up95}), which
suggest that the appearance of an AGN strongly depends on the orientation.
However, it is not well understood if the jet velocity is the same in
all radio sources,  if it is correlated to other source properties as
the radio power, and if velocity structures are common.\\
In this review I will discuss the
jet  kinematic properties.
 
\section{The Measurement of the Jet Velocity}
\subsection{Proper Motion}

Many AGNs contain compact radio sources with different components which
appear to move apart. Multi epoch studies of these sources allow a direct 
measure of the apparent jet pattern velocity ($\beta_a$). The observed 
distribution of the apparent velocity shows a large range of values 
\cite{vc94}; \cite{k2000}. 
From the measure of $\beta_a$ we can 
derive constraints on $\beta_p$ and $\theta$ where $\beta_p$ is the intrinsic
velocity of the pattern flow and $\theta$ is the jet orientation with respect
to the line of sight:

$\beta_p$ = $\beta_a/(\beta_a cos\theta + sin\theta)$

A main problem is to understand the difference between the bulk and pattern
velocity. In few cases where proper motion is well defined and the bulk 
velocity is strongly constrained, there is a general agreement between the 
pattern velocity and the bulk velocity (see e.g. NGC 315, \cite{cot99},
and 1144+35, here). However, in the same source we can have different
pattern velocities as well as stationary and high velocity moving structures.
Moreover, we note that in many well
studied sources the jet shows a smooth and uniform surface brightness and
no (or very small) proper motion (as in the case of Mkn 501,
Giroletti et al. in preparation, and M87 in the region at $\sim$ 1 pc from the
core \cite{jun99}).

\subsection{Jet Sidedness}

Assuming that the jets are intrinsically symmetric we can use the observed
jet to counter-jet brightness ratio R to constrain the jet 
bulk velocity $\beta$c
and its orientation with respect to the line of sight ($\theta$):

R = (1+$\beta cos\theta)^{2+\alpha}$ (1-$\beta cos\theta)^{-(2+\alpha)}$

where $\alpha$ is the jet spectral index (S($\nu$) $\propto$ $\nu^{-\alpha}$).
Some problems can be related to this measurement:
free free absorption may affect the observed jet brightness, moreover,
we cannot derive strong constraints in intrinsically low luminosity jets:
e.g. in 3C264 where a well studied optical and radio jet is present
\cite{lar99} the highest j/cj ratio is R $>$ 37 which implies 
that the source is
oriented at $\theta$ $<$ 52$^\circ$ with a jet moving with $\beta$ $>$ 0.62.

\subsection{Radio Core Dominance}

The core radio emission measured at arcsecond resolution is dominated
by the Doppler-boosted pc-scale relativistic jet.  
The source radio power measured at low frequency (e.g. 408 MHz), instead, 
is due 
to the extended emission, which is not affected by Doppler 
boosting. At low frequency the observed core radio emission is
not relevant since it is mostly self-absorbed. 
Given the existence of a general correlation between the core and total
radio power discussed in  \cite{gg01},
we can
derive the expected intrinsic core radio power from the {\it unboosted}
total radio power.
The comparison between the expected intrinsic core radio
power and the observed core radio power will give constraints on the jet 
velocity and orientation \cite{gg01}.
We note that the core radio power is best measured at 5 GHz where
it is dominant because of the steep spectrum of the extended emission, 
self-absorption is not relevant, and high angular
resolution images allow us to separate the core from the extended jet
emission.\\
Plotting the observed core radio power at 5 GHz against the total radio power 
at 408 MHz \cite{gg88} we find a large dispersion in the core 
radio power. This is expected because
of the strong dependance of the observed core radio power on $\theta$ and 
$\beta$.
From the data dispersion, assuming that no selection effect is present in 
the source orientation ($\theta$ = 0$^\circ$ to 90$^\circ$), 
we can derive that the 
jet Lorentz factor $\gamma$
has to be $<$ 10 otherwise we should observe a larger core radio power 
dispersion.\\
A problem related with this measurement is a possible nuclear intrinsic 
variability. However, we note that a flux density variability larger than a 
factor 2 in the nuclear radio emission is uncommon.

\subsection{Other Methods}

{\it Synchrotron-Self-Compton Emission}: in principle when the core angular 
size is known, the comparison between the
observed and predicted nuclear non thermal X-ray emission gives a limit to
the source Doppler factor $\delta$ \cite{ghis93}.
A problem in this approach is the too large uncertainty in the observed
nuclear X-Ray emission and in the core angular size.

{\it Arm Length Ratio}:
by comparison of the size of the approaching (L$_a$) and 
receding jet (L$_r$) we derive:
 
L$_a$/L$_r$ = (1 + $\beta$ cos$\theta$)(1 - $\beta$ cos$\theta$)$^{-1}$
 
A major problem is that
the source size is related to the uv coverage and image sensitivity, moreover,
the jet interaction with the ISM can affect the source size.

{\it Brightness Temperature}: 
estimates of the T$_b$ from source variability or high resolution images
(e.g. Space VLBI) can give
constraints on the source
Doppler factor (up to 1000 for Intra Day Variable (IDV) sources). I will not 
discuss this point,
since IDV is still a hot and controversial topic (see
\cite{cim02}).\\
Readhead \cite{read94} suggested an alternative method:
 
$\delta$ = T$_{b,VLBI}/T_{eq}$
 
where the brightness temperature under equipartition conditions is 
T$_{eq}$ $\sim$ 5 10$^{10}$ K.

\section{Results}

In the literature many single sources and/or samples have been studied.
In most cases observational data are in agreement with Unified Scheme Models
and with the presence of a high velocity pc scale jet.\\
To derive statistical properties of radio jets on the pc scale, we 
extended the results discussed in \cite{gg01}
selecting all sources of the 3CR and B2 catalogues in a sky region,
with z $<$ 0.1 (to avoid selection effects on the source orientation
and core radio power).\\
The sample consists of 95 sources, 53 of which have been observed with
VLBI. Among them we have 32 FR I, 10 FR II, 10 compact flat spectrum
sources (2 BL-Lacs, 1 CSO), and 1 CSS.
From these preliminary results (Giovannini et al., in preparation),
we found that in all sources pc scale jets move at high velocity. No
correlation has been found between the jet velocity and  the core or total 
radio power. 
Highly relativistic parsec scale jets
are present regardless of the radio source power. Sources with a different kpc 
scale morphology, and
total radio power have pc scale jets moving at similar velocities.\\
We used the estimated jet $\beta$ and $\theta$ to derive
the Doppler factor $\delta$ for each source,
and the corresponding intrinsic core radio power (assuming $\alpha$ = 0):
 
P$_{c-observed}$ = P$_{c-intrinsic}$ $\times$ $\delta^2$

\begin{figure}
\centerline{\epsfig{figure=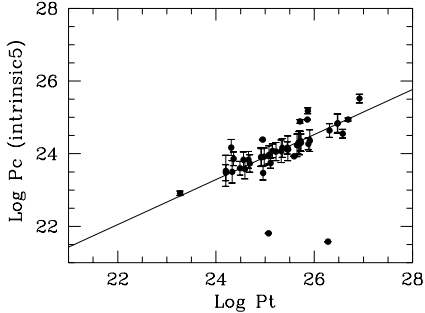,height=4.0truecm}}
\caption{Total radio power at 408 MHz versus the intrinsic core
radio power at 5 GHz derived with $\gamma$ = 5.
The continuum line represents the general correlation between the core and 
the total
radio power (see \cite{gg01}).}
\label{figopt1}
\end{figure}

We found a good correlation between  P$_{c-intrinsic}$ and P$_{t}$ with a 
small dispersion
since plotting P$_{c-intrinsic}$, we removed the spread due to
the different orientation angles (Fig. 1).
We found that a Lorentz factor $\gamma$ in the
range 3 to 10 is consistent with the observational data.
The Lorentz factor cannot be $>$ 10 for the previous considerations 
(Sect. 2.3).
It cannot be $<$ 3 to remove the dispersion due to the different 
source orientations.
The result that sources with different P$_{t}$ and kpc scale morphology
show the same correlation, implies that {\bf all pc scale
jets have
a similar velocity}.
Two sources do not follow the general correlation: M87 where
we should
have a higher jet velocity to fit with the general correlation and
3C 192 which core radio power is lower than expected (it could be in a 
pre-relic phase).
 
\section{Acceleration and Deceleration in Jets}

In some sources evidences of increasing velocity in pc scale jets 
have been found: e.g. M87 \cite{bir95}; 
3C84 \cite{dha98}; Cygnus A \cite{kr98}.
However, in these sources the jet velocity was measured from the jet
apparent motion. Thus the increasing velocity could be
non intrinsic but due to a change in the jet direction
or to a change in the jet pattern velocity unrelated
to the jet bulk velocity. In NGC 315 Cotton et al. \cite{cot99} found 
an increasing
jet velocity in the 5 inner pc from the core both from proper motion 
measurements AND
from the sidedness ratio.\\
A jet deceleration is evident in the FR I radio sources (see Laing this 
volume). 
In many low power sources there are 
observational evidences that jets are relativistics at their beginning and
strongly decelerate within a few kpc from the core (see e.g. 3C449 
\cite{fer99}).\\
In FR II radio sources kpc scale jets are still affected by Doppler
boosting effects, but the jet sidedness ratio decreases with the 
core distance
implying a velocity decrease. Observational data are consistent with $\gamma$
$\sim$ 2 on the kpc scale \cite{br94}.

\section{Velocity Structures}

An evident limb-brightened jet morphology on the pc scale is present in some
FR I sources as 1144+35, Mkn 501, 3C264, M87, 0331+39 \cite{gg01},
and also in a few high power radio sources as 1055+018 \cite{att99}, 
3C 236 (Giovannini et al. in preparation). We interpret the
limb-brightened structure as due to a different Doppler boosting effect in 
a two-velocity relativistic jet. If the source is oriented at a relatively 
large angle with respect to the line of sight, the inner very high 
velocity spine
could be strongly deboosted, while the slower external layer could be boosted
and become brighter than the inner jet region. Therefore only sources in
a small range of $\theta$ will appear limb-brightened. For this reason, and
for the observational 
difficulties of transversally resolving the radio jets, 
we expect that the number of sources exhibiting limb-brightened jets is low,
as observed.

At present it is not clear if the velocity structure is strictly related
to the jet interaction with the ISM as suggested by \cite{gg01}
or it is an intrinsic jet property. The presence of this structure very near to
the central core as in M87 \cite{jun99} and MKn 501 (Giroletti et
al. in preparation) is in favour of a jet intrinsic property 
in the inner region (as discussed
by Meier in this volume), 
whereas a jet velocity decrease
due to the ISM is likely to be present in an intermediate region (from the
pc to the kpc scale) \cite{gg01}.

\section{The Source 1144+35}

\begin{figure}
\centerline{\epsfig{figure=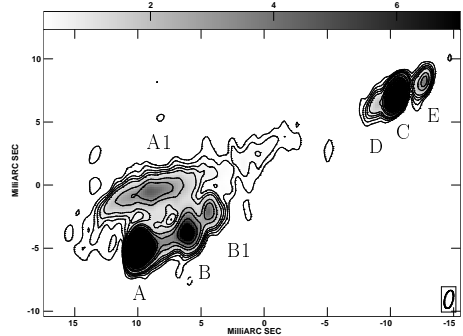,height=4.5truecm}}
\caption{VLBI Image of 1144+35 at 8.4 GHz. Labels are: A,A1,B, and B1 jet 
substructures moving with $\beta_a$ = 2.7; D the jet beginning, C the core,
E the counter-jet.}
\end{figure}

1144+35 is a giant radio galaxy which shows a strong extended
jet and a short counter-jet on the pc scale \cite{gg99}.
The main jet is limb-brightened with evident substructures in the external
shear layer which are 
moving with $\beta_a$ $\sim$ 2.7 (Fig. 2). 
New recent observations at 8.4 GHz have allowed to measure a possible
proper motion also in the cj: $\beta_{a-cj}$ = 0.3 $\pm$ 0.1 in the
time range 1995 - 2002 with 4 different epochs (Giovannini et al. in 
preparation). 
The jet bulk velocity 
is in the range 0.8c - 0.9c, as derived from the jet sidedness ratio, for the 
bright jet 
external regions. The pattern
velocity, derived from the main jet proper motion in the same regions,
is $>$ 0.8c. In both cases
the derived jet orientation angle to the line of sight is $\theta$ $<$ 
37$^\circ$.
Moreover, according to \cite{mir94} using both the jet 
and counter jet
proper motions we can obtain directly $\theta$ = 30$^\circ$ and
$\beta$ = 0.9. 
From these data we can obtain a comprehensive scenario, as follows:
1) the bulk and pattern velocity are the same;
2) the shear layer Doppler factor is $\sim$ 2 (being $\gamma$ $\sim$ 2.3)
while assuming $\gamma$ = 15
for the inner spine, its Doppler factor is 0.7 in agreement with
the observed brightness distribution;
3) if the external shear layer started with the same velocity of the inner 
spine,
its velocity decreased from 0.998c to 0.9c in less than 100 pc suggesting
an intrinsic origin for the jet velocity structure (Meier this volume).

\section{Conclusions}

The jet velocity and orientation can be estimated with good observational
data. In a few sources the bulk and pattern jet velocity are the same, but
the problem is still open.\\
The pc scale jet velocity is highly relativistic ($\gamma$ = 3 to 10) in 
both high
and low power radio sources, with no relation with the total radio
power and the large scale morphology.\\
The jet velocity from the pc to the kpc scale decreases dramaticaly in FR I 
radio sources and
more slowly in FR II sources. In a few cases a jet acceleration has been
found in the source inner regions (few pc).\\
A two velocity regime has been found in some low power sources and in a few
high power sources. The different Doppler factor can explain the observed
limb-brightened structures visible in some high resolution images.
The presence of a shear layer could be an intrinsic property;
a velocity decrease in the shear layer could be due to the
interaction with the surrounding ISM in the tens of pc -- kpc scale.

\end{document}